\documentclass[letterpaper, 10 pt, conference]{ieeeconf}  

\IEEEoverridecommandlockouts                              

\usepackage{cite}

\usepackage{graphicx}

\usepackage{amsmath}

\usepackage{algorithmic}

\usepackage[caption=false,font=footnotesize]{subfig}

\usepackage{placeins}

\usepackage{url}

\usepackage{graphicx} 

\usepackage{amssymb} 	
\usepackage{bm} 		
\usepackage{siunitx}	
\usepackage{booktabs} 	
\usepackage{multirow}   
\usepackage{colortbl}
\usepackage{array}
\usepackage{lmodern}
\definecolor{lightgrey}{gray}{0.9}

\title{\LARGE \bf
A Method for Classifying Test Bench Configurations in a Scenario-Based Test Approach for Automated Vehicles*
}

\author{Markus Steimle$^{1}$, Till Menzel$^{1}$, and Markus Maurer$^{1}$
\thanks{*This work was supported by the project PEGASUS funded by the German Federal Ministry for Economic Affairs and Energy.}
\thanks{$^{1}$All authors are with the Institute of Control Engineering,
        Technische Universit\"at Braunschweig, 38106 Braunschweig, Germany
        {\tt\small \{steimle,menzel,maurer\}@ifr.ing.tu-bs.de}}%
}

\begin{document}

\maketitle
\thispagestyle{empty}
\pagestyle{empty}

%
\begin{abstract}%
	The introduction of automated vehicles demands a way to prove their safe operation.
However, validating the safety of automated vehicles is still an unsolved problem. 
While the scenario-based test approach seems to provide a possible solution, it requires the execution of a high amount of test cases.
Several test benches, from actual test vehicles to partly or fully simulated environments, are available, but choosing the optimal test bench, e.g. in terms of required execution time and costs, is a difficult task.
Every test bench provides different elements, e.g.~simulation models which can be used for test case execution.	
The composition of elements at a specific test bench is called test bench configuration.
This test bench configuration determines the actual performance of a test bench and, therefore, whether the run of a particular test case provides valid test case results with respect to the intended purpose, e.g.~a safety validation.
For an effective and efficient test case execution, a method is required to assign test cases to the most appropriate test bench configuration.
Therefore, it is indispensable to have a method to classify test bench configurations in a clear and reproducible manner.
In this paper, we propose a method for classifying test benches and test bench configurations and illustrate the classification method with some examples.
The classification method serves as a basis for a systematic assignment of test cases to test bench configurations which allows for an effective and efficient test case execution.%
\end{abstract}%

\section{Introduction}
\label{sec_introduction}

Validating that the vehicle behaves safely is a challenge in the introduction of automated vehicles with an SAE Level $\geq \ $3~\cite{SAE.2016}.      
A conventional, distance-based validation as used for driver assistance systems is not applicable for reasons of time and cost~\cite{Wachenfeld.2016b}; however, the scenario-based test approach seems to offer a possible solution~\cite{Schuldt.2017}.
Bagschik~et~al.~\cite{Bagschik.2018b} and Menzel~et~al.~\cite{Menzel.2018b} propose a method for deriving scenarios during the development process, which then can be used to derive test cases.
Applying this method results in a high amount of test cases, which need to be executed.
For the execution of a test case, there are several test benches available, from actual test vehicles to partly or fully simulated environments.
Choosing the optimal test bench is a challenging task.
There are several reasons to use simulative test methods, if applicable, instead of actual driving tests: e.g.~test case execution can be faster than real time, the developed simulation models can be used several times, there are reproducible test conditions, the risk to people and material is greatly reduced when testing safety-relevant driving functions, and it might be possible to reduce costs in the development process.
Therefore, test cases should be executed with simulative test methods whenever possible.

Whether a test case can be executed with a specific test bench depends on the properties of this test bench.
Every test bench provides different elements, partially with the same purpose, e.g.~different vehicle dynamics simulation models with different accuracies and execution times.
The composition of these elements determines the actual performance of the test bench and therefore whether the run of a particular test case with this test bench configuration provides valid test case results.
Valid test case results are test case results that can be used for the intended purpose, for example a safety validation.
Only if validated simulation models with respect to the intended purpose are used for the run of a test case can valid test case results be generated with the corresponding test bench configuration. 
Only then is a valid statement, e.g.~for a safety validation, possible.

Currently, in a distance-based test approach, as used for driver assistance systems, test cases are mostly executed in form of driving tests with various test vehicles.
When simulation is used for test case execution, experts decide manually which simulation models and, thus, which test bench configuration should be used.
Therefore, experts decide which test bench configurations may be valid.
In order to check this validity after test case execution, it is still common that experts manually compare graphs from simulation data and measurement data from driving tests~\cite{Viehof.2018b}.
Within the scope of safety validation of automated vehicles, the significance is limited with this manual and subjective expert-based method.

Especially, because of the large number of test cases and different elements provided at test benches it is no longer feasible to manually select test bench configurations for each test case in order to generate valid test case results effectively and efficiently.
For efficient and effective execution of the large number of test cases, an automated assignment of test cases to the most appropriate test bench configuration is required.
For that, a method for an unambiguous classification of test benches and test bench configurations is indispensable in order to formalize the properties of a test bench and in particular the properties of the possible test bench configurations.

There are generic approaches to classify test methods (e.g.~\cite{Schuldt.2015b}) and test benches (e.g.~\cite{Strasser.2012} or~\cite{NeumannCosel.2014}).
However, these generic classification methods do not take into account the specific elements (e.g.~simulation models) which are available at a specific test bench.
For the classification of test benches and test bench configurations, the consideration of the elements provided at a specific test bench is absolutely necessary.
Therefore, an unambiguous classification of test bench configurations is not possible with these methods.

In this paper, we propose a method to unambiguously classify test benches and test bench configurations. 
For this, we systematically derive dimensions to classify all the functionalities that a test bench needs to provide in order to execute a test case.
The dimensions are identified by an analysis of a functional system architecture of an automated vehicle and its interactions with the driver or user and the environment.  
This classification method serves as the basis for a systematic assignment of test cases to test bench configurations in order to execute test cases efficiently and effectively.

This paper is structured as follows: 
Section~II gives a short motivation based on selected related work regarding classification methods for test methods and test benches.
Section~III presents the proposed method for classifying test benches and test bench configurations. 
Afterwards, in section~IV the proposed classification method is evaluated by showing various examples of test benches and test bench configurations.
Finally, section~V gives a conclusion and addresses future work.
\section{Related Work}
\label{sec_related_work}

Studies have classified various test methods and test benches by their components, but the components used for this classification vary. 
In this section, three classification methods, including the distinguished test methods or test benches, are reviewed.

\begin{figure}[ht]
	\centering
	\includegraphics[width=0.7\columnwidth]{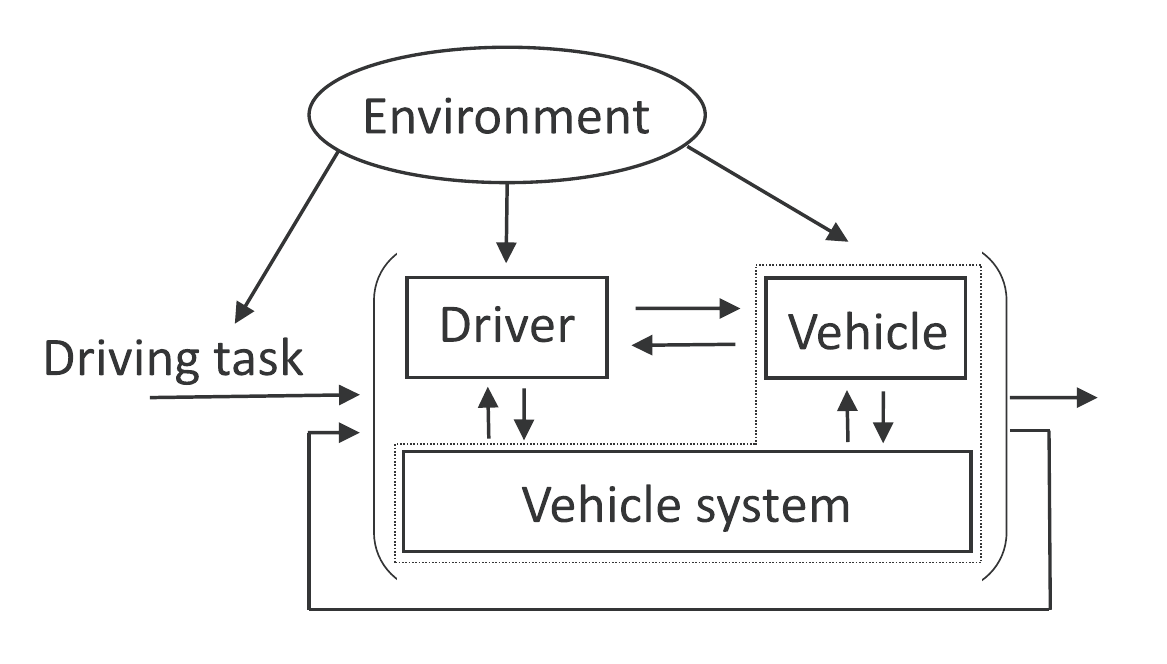}
	\caption{Structure of a human machine system, based on Bubb \& Schmidtke~\cite{Bubb.1993}, translated from Strasser~\cite{Strasser.2012}}
	\label{fig_Struktur_Strasser}
\end{figure}

Strasser~\cite{Strasser.2012} classifies test benches based on the structure of a human machine system, as shown in Fig.~\ref{fig_Struktur_Strasser}. 
It consists of four modules: driver, environment, vehicle, and electric vehicle system that represents the test object. 
Each module can be simulated or real.
Due to the combination of simulated and real modules, Strasser~\cite{Strasser.2012} classifies the test benches software-in-the-loop, hardware-in-the-loop, driver-in-the-loop, vehicle-in-the-loop, and test vehicles (onboard test or rapid prototyping).
However, this binary classification of simulated and real modules is not able to clearly classify every element of a test bench. 
As an example Schuldt~et~al.~\cite{Schuldt.2015b} mention a balloon vehicle, which is neither a real nor a simulated vehicle, but is emulated by a hardware with similar dimensions. 
This distinction may be significant for testing sensors because sensor data representing the balloon vehicle may be different from the sensor data of a real or a simulated vehicle.
For this reason, based on Wachenfeld \& Winner~\cite{Wachenfeld.2016b} Schuldt~et~al.~\cite{Schuldt.2015b} suggest the classification of emulated components between real and simulated ones.

\begin{figure}[ht]
	\centering
	\includegraphics[width=0.99\columnwidth]{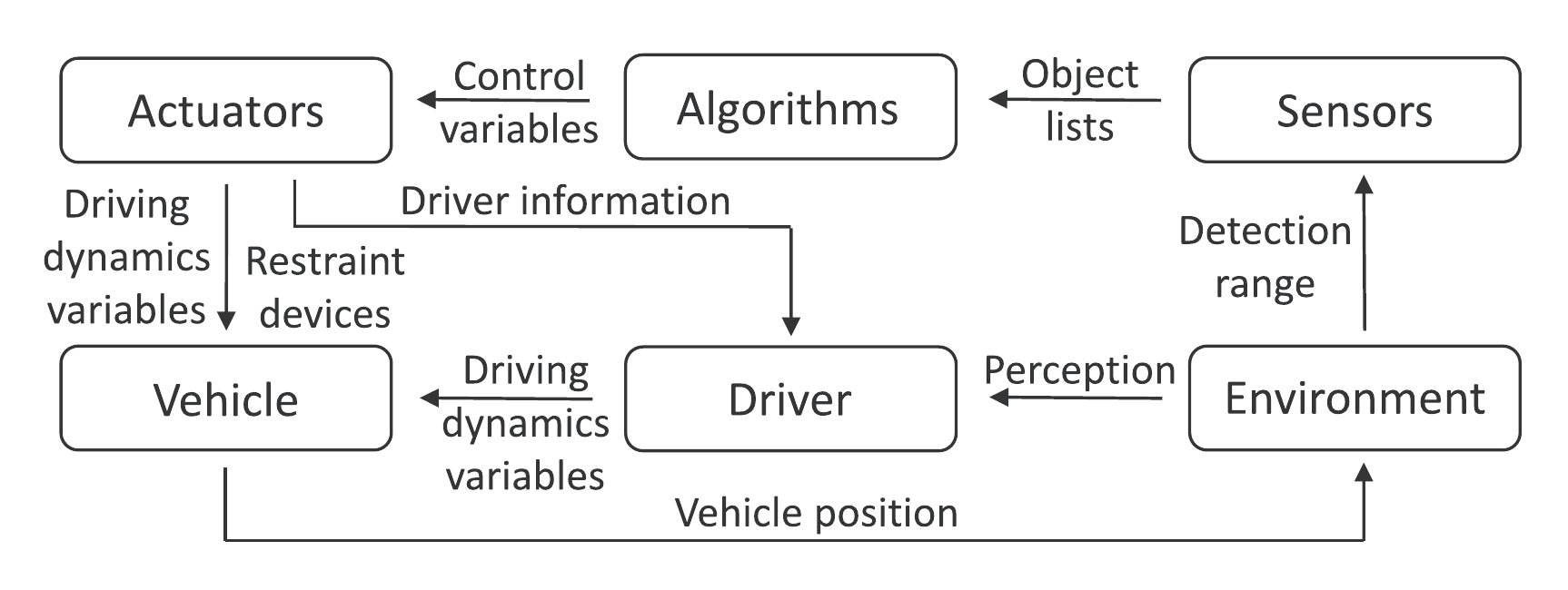}
	\caption{Interaction of the driving function with the driver, vehicle, and environment, translated from von Neumann-Cosel~\cite{NeumannCosel.2014}}
	\label{fig_Struktur_Neumann}
\end{figure}

Von Neumann-Cosel~\cite{NeumannCosel.2014} classifies test benches based on the interaction of the driving function (a combination of sensors, algorithms, and actuators) with the driver, the vehicle, and the environment, as shown in Fig.~\ref{fig_Struktur_Neumann}.
Each component can be simulated or real.
Due to the combination of simulated and real components, von Neumann-Cosel~\cite{NeumannCosel.2014} classifies the test benches concept-in-the-loop, software-in-the-loop, hardware-in-the-loop, driver-in-the-loop, vehicle-in-the-loop, and a test vehicle with a driving robot, which operates the steering wheel and pedals and thereby replaces the driver.

\begin{figure}[ht]
	\centering
	\includegraphics[width=0.78\columnwidth]{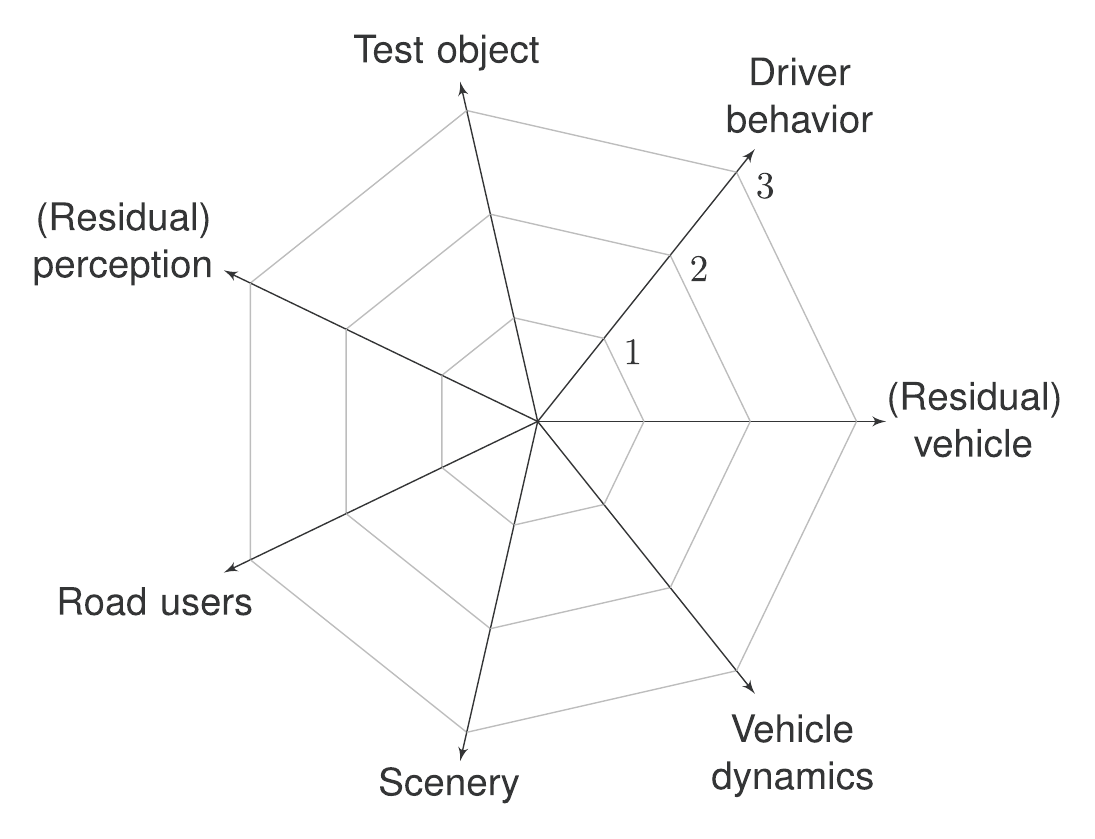}
	\caption{Empty radar chart with the stages: 1~=~simulated, 2~=~emulated, 3~=~real, translated from Schuldt~\cite{Schuldt.2017}}
	\label{fig_KV_empty}
\end{figure}

Based on Schuldt~et~al.~\cite{Schuldt.2015b} Schuldt~\cite{Schuldt.2017} classifies test methods in different dimensions: test object, driver behavior, (residual) vehicle, vehicle dynamics, (residual) perception, road users, and scenery.
Each dimension can be simulated, emulated or real.
With the combination of these dimensions Schuldt~\cite{Schuldt.2017} classifies the test methods software-in-the-loop, hardware-in-the-loop, driver-in-the-loop, vehicle-hardware-in-the-loop, and vehicle-in-the-loop.
To visualize the results, Schuldt~\cite{Schuldt.2017} uses radar charts, where every spoke represents a dimension of the test method.
An empty radar chart is shown in Fig.~\ref{fig_KV_empty}.

The authors of the mentioned publications classify test benches and test methods in a generic way. 
So far, we have not found any publications which propose methods to classify specific test bench configurations.
This is indispensable for an automated assignment of test cases to test bench configurations.
For this reason, we propose a method for the classification of test benches and test bench configurations in the next section.

\section{Classification method for test benches and test bench configurations}
\label{sec_classification_method}

The proposed classification method is based on functionalities which need to be provided at a test bench to execute a test case.
To derive these functionalities in a systematic way, we examine how an automated vehicle interacts with its environment and a driver or a user within a scenario.
First, we take a look at a test bench where all elements are real (test vehicle).
Secondly, we take a look at a test bench where all elements are simulated (software-in-the-loop test bench).
Thus, we examine the two extreme characteristics of test benches. 

Steimle~et~al.~\cite{Steimle.2018a} describe that a test case consists of a scenario and evaluation criteria. 
According to Bagschik~et~al.~\cite{Bagschik.2018}, based on Schuldt~\cite{Schuldt.2017}, a scenario can be structured by the 5-layer model. 
Therefore, a scenario consists of the road-level, traffic infrastructure, temporary manipulation of road-level and traffic infrastructure, movable objects, and environment conditions.

A distinction between driver and user is considered meaningful since there is only a user and not a driver in automated mode, who interacts with the vehicle.
Depending on the automation level of the vehicle, the user can take over control by driving him- or herself. 

\begin{figure}[ht]
	\centering
	\includegraphics[width=0.99\columnwidth]{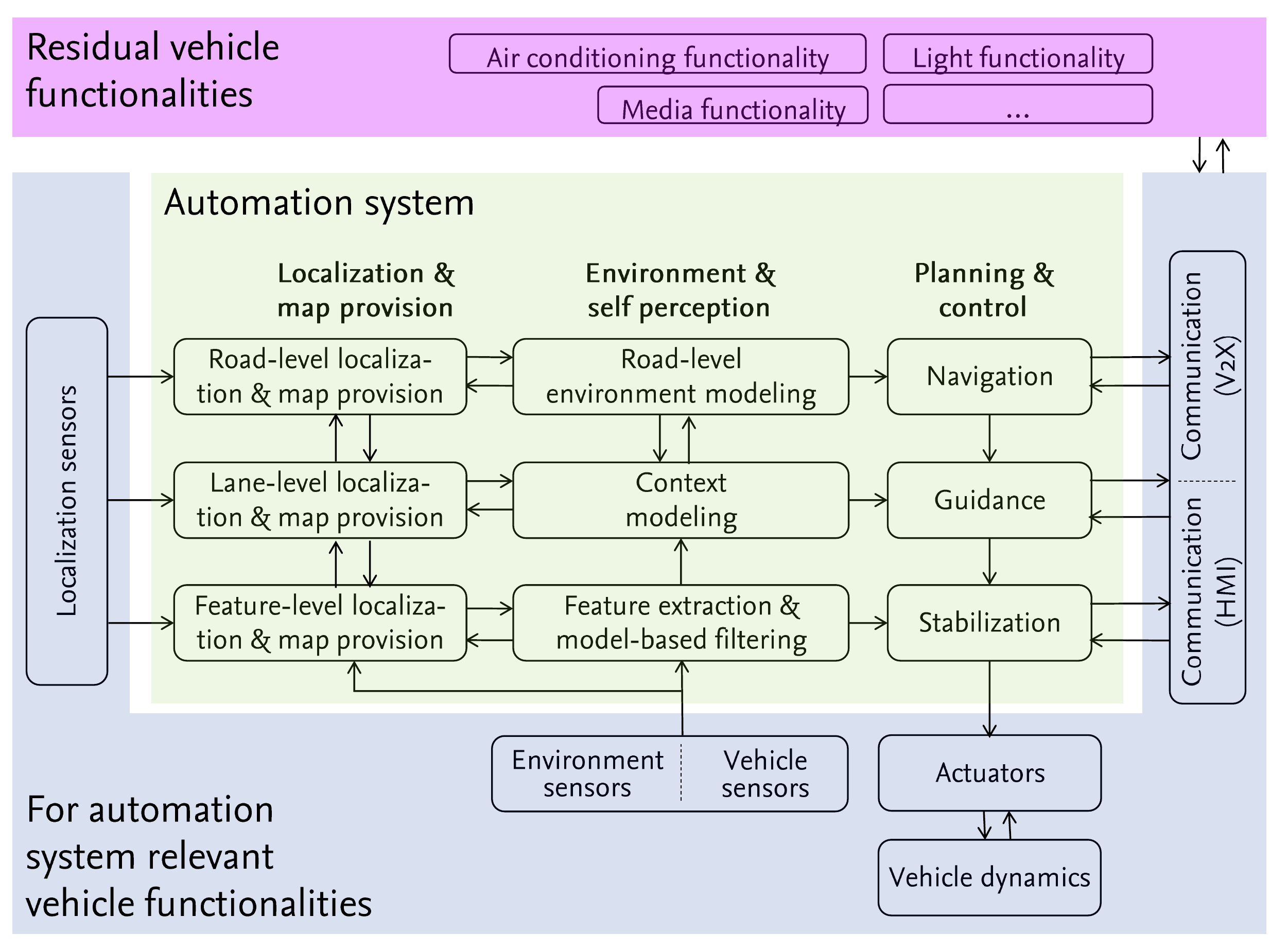}
	\caption{Functional system architecture of an automated vehicle, based on Matthaei \& Maurer~\cite{Matthaei.2015}}
	\label{fig_general_architecture}
\end{figure}

The functionalities of an automated vehicle are described with a functional system architecture developed at the Institute of Control Engineering of the TU Braunschweig.
This architecture was repeatedly discussed and evolved (e.g.~in~\cite{Matthaei.2015},~\cite{Matthaei.2015b} or~\cite{Ulbrich.2017}). 
For the complete functional description of an automated vehicle, this architecture requires two additional functional blocks: ``vehicle dynamics'' and ``residual vehicle functionalities''.
The ``vehicle dynamics'' describes the movement of an automated vehicle in the environment due to the actuation of the actuators.
The ``residual vehicle functionalities'' describe functionalities of the components which are not directly required for operating the automation system, but are necessary for the complete functional description of an automated vehicle.
Fig.~\ref{fig_general_architecture} shows the resulting functional system architecture of an automated vehicle.
The automation system and the vehicle functionalities are color coded and named. 
The residual vehicle functionalities can influence the functional blocks of the vehicle functionalities which are relevant for the automation system and vice versa. 
This can indirectly influence the automation system. 
For clarity, these connections are only drawn to the vehicle functionalities which are relevant for the automation system and not to each individual functional block.
For a more detailed overview and explanation of the other functional blocks shown in Fig.~\ref{fig_general_architecture}, see~\cite{Matthaei.2015} or~\cite{Ulbrich.2017}.

An automated test vehicle moves in a real environment and may be controlled by a real driver or user. 
This means, the interfaces of the functional system architecture shown in Fig.~\ref{fig_general_architecture}, which describes the automated vehicle, communicate with the real environment and the real driver or user.
To operate the automation system at a software-in-the-loop test bench, the interfaces communicate with a simulated environment and a simulated driver or user. 
Consequently, a simulation environment must provide the same data for the interfaces of the automation system. 

Based on these two extreme characteristics, the functionalities of a test bench can be described in a structured manner. 
Radar charts are suitable for a clear presentation of this multi-dimensional description. 
Each dimension of the radar chart represents a specific functionality of a test bench. 
Each dimension may be real, emulated or simulated. 
Therefore, we use a nominal scale with the discretization stages real, emulated, and simulated to subdivide each dimension.
The proposed definitions of these three stages are listed below:

\begin{itemize}
	\item \textbf{Real}: If an element is real, it will be used as planned in the target vehicle.
	\item \textbf{Emulated}: If an element is emulated, a comparable element with respect to the target hardware that behaves like the real element is used.
	At the interfaces the emulated element shows identical functionality and behavior.~\cite{Schuldt.2017}
	\item \textbf{Simulated}: If an element is simulated, a simulation model which shows identical functionality and behavior is used.
	The simulation model has no hardware reference and is implemented exclusively in software.~\cite{Schuldt.2017}
\end{itemize}

These three stages and definitions reflect our current state of research.
The question remains as to whether the discretization of the dimensions with three stages is sufficient for a test case assignment method. 
A further subdivision of the stages may be required to allow a more accurate classification of the elements, for example, to be able to distinguish different development stages of a real element.
However, this more detailed discretization does not argue against the proposed classification method, but extends it.
Further investigations are necessary in the development of the test case assignment method.

\begin{figure}[ht]
	\centering
	\includegraphics[width=0.94\columnwidth]{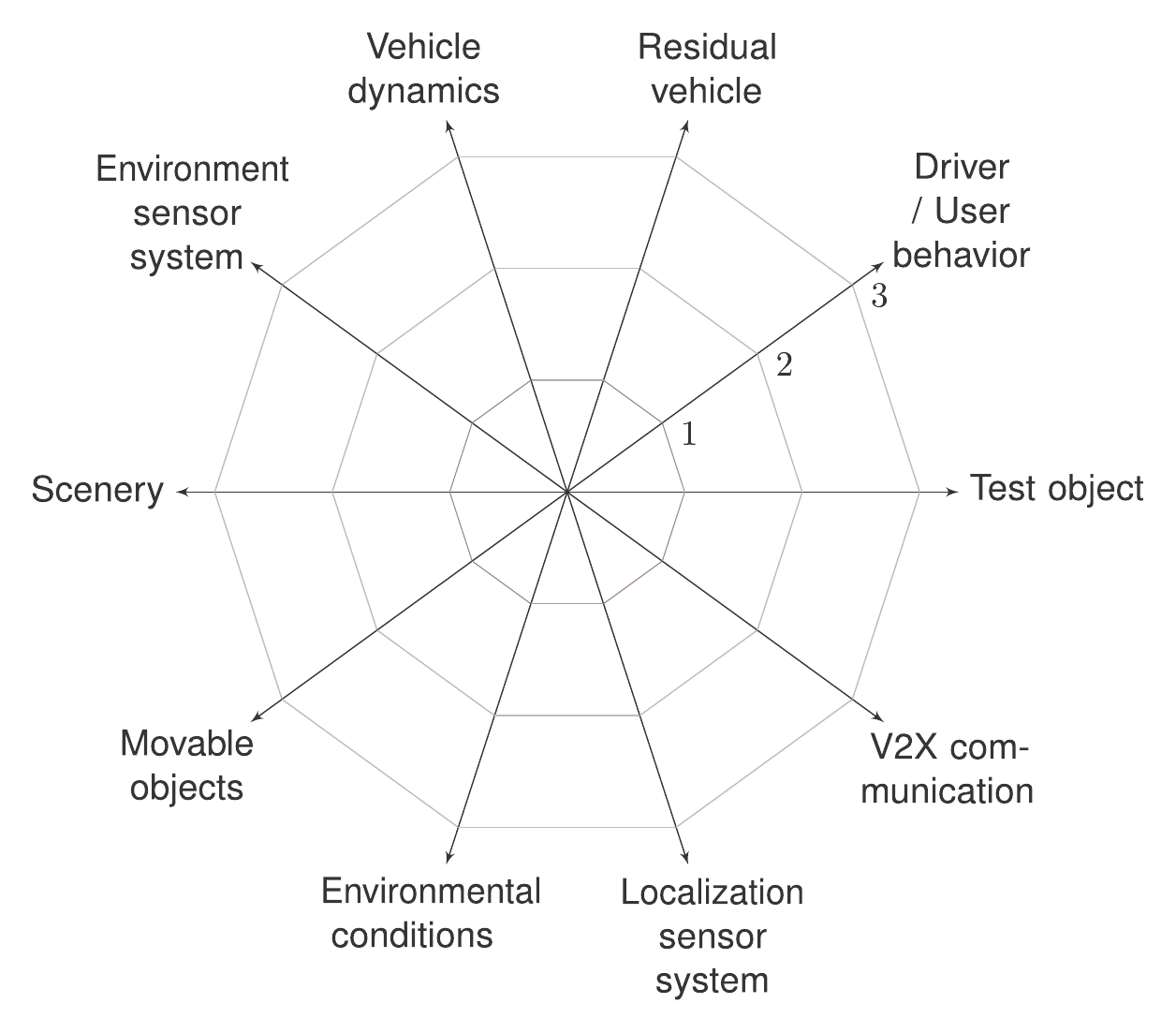}
	\vspace{-0.3cm}
	\caption{Empty radar chart with the stages: 1~=~simulated, 2~=~emulated and 3~=~real}
	\label{fig_general_spider}
\end{figure}

The derived dimensions are used to classify test benches and test bench configurations.
Thereby, the test methods are classified as well.
Fig.~\ref{fig_general_spider} shows the derived dimensions in a radar chart. 
At a test bench, each functionality is implemented by one or more elements, e.g.~for the dimension environment sensor system, a real sensor or a sensor simulation model.
Therefore, to clearly classify test benches and test bench configurations, we have to consider the specific elements provided at a test bench.
Examples of real, emulated, and simulated elements at each dimension are listed in Tab.~\ref{tab_dimensions}.
The derived dimensions with proposed definitions are listed below.

\begin{table*}[h]
	\caption{Examples of real, emulated, and simulated elements at each proposed dimension}
	\label{tab_dimensions}
	\begin{center}
		\begin{tabular}{llll}
			\toprule
			\textbf{Dimension}							& \textbf{Real element} 		& \textbf{Emulated element} 					& \textbf{Simulated element} \\
			\midrule
			Test object 								& series control unit 			& rapid prototyping control unit				& program code in development software  \\
			
			\rowcolor{lightgrey}	Driver / user 		& test driver 					& driving robot (that moves the steering 		& driver simulation model	\\		
			\rowcolor{lightgrey}	behavior 			&  		    					& wheel)  										&  	\\		
			
			Vehicle  									& vehicle dynamics of 			& vehicle dynamics of another vehicle which		& vehicle dynamics simulation model that   \\
			dynamics 									& the series vehicle 			& behaves identical to the series vehicle  		& calculates the movement of the vehicle \\	
			
			\rowcolor{lightgrey}	Environment 		& series radar sensor   		& pre-series radar sensor 						& radar simulation model \\
			\rowcolor{lightgrey}	sensor system		& 								& 												&  \\	
			
			Scenery 									& real curb or real tree 		& artificial curb or artificial tree			& simulation model of a curb or tree  \\		
			
			\rowcolor{lightgrey}	Movable  			& series vehicle  				& balloon vehicle or crash target  				& vehicle simulation model  \\	
			\rowcolor{lightgrey}	objects 		 	&                				&    											& \\
				
			Environmental								& real rain or 	 				& wetted lane or artificially 					& rain simulation model \\
			conditions 									& real fog 						& generated rain  								&   \\			
			
			\rowcolor{lightgrey}	Localization  	 	& series localization    		& development localization sensor system 		& localization simulation model \\
			\rowcolor{lightgrey}	sensor system 		& sensor system  				&   											&  \\				
			
			Vehicle-to-X (V2X)							& V2X-data generated     		& V2X-data generated by replicated 				& V2X-data generated by   \\
			communication								& by an other vehicle 			& transmitter like prototypical hardware		& a V2X-simulation model   \\		  
			
			\rowcolor{lightgrey}	Residual  			& series vehicle is				& similar vehicle or similar hardware			& rest-bus simulation model \\
			\rowcolor{lightgrey}	vehicle    			& used to operate 				& is used to operate the test object			& is used to operate the test object \\		
			\rowcolor{lightgrey}				  		& the test object				& 												&  \\		
			\bottomrule
		\end{tabular}
	\end{center}
\end{table*}

\begin{itemize}
	\item \textbf{Test object}: The test object describes the system components to be tested, including the execution platform~\cite{Schuldt.2017}.

	\item \textbf{Driver / User behavior}: 	The driver / user behavior describes the behavior of the driver or the user and the operation of the interfaces of the vehicle~\cite{Schuldt.2017}.

	\item \textbf{Vehicle dynamics}: The vehicle dynamics describe the movement of the vehicle~\cite{Schuldt.2017}.

	\item \textbf{Environment sensor system}: The environment sensor system describes the characteristics of the environment sensors~\cite{Schuldt.2017}.

	\item \textbf{Scenery}: The scenery subsumes all geo-spatially stationary elements of the environment~\cite{Ulbrich.2015b}.

	\item \textbf{Movable objects}: Movable objects describes objects that are moving through kinetic energy or have the ability to move through energy or abilities~\cite{Ulbrich.2015b}.

	\item \textbf{Environmental conditions}: Environmental conditions are external influences, such as weather and lighting conditions~\cite{Bagschik.2018}.

	\item \textbf{Localization sensor system}: The localization sensor system describes the characteristics of the sensors used for localization.

	\item \textbf{V2X communication}: The V2X communication describes wireless communication with other elements.

	\item \textbf{Residual vehicle}: The residual vehicle describes the elements of the vehicle which, in addition to the elements of the remaining dimensions, are additionally required for the operation of the test object.
	The residual vehicle does not have to cover all vehicle functionalities, but only the parts that are needed for the operation of the test object.
	However, the residual vehicle can also cover other vehicle functionalities that are not required for the operation of the test object.

\end{itemize}

A test bench can have various elements in one dimension and stage. 
For example, in the dimension vehicle dynamics at the stage simulated, there might exist the elements single-track simulation model and double-track simulation model.

Each element has certain characteristics, such as validity, costs per operating time, and complexity in terms of necessary execution time.

In all dimensions a combination of simulated, emulated, and real elements of the respective dimension might exist. 
Thus, in a test case execution with a vehicle-in-the-loop test bench, real vehicles might exist as movable objects and additionally emulated vehicles might exist as movable objects when balloon vehicles are used. 
Both are perceived by the present environment sensors. 
In addition, simulated vehicles might exist as movable objects, which are perceived by sensor models. 
A detailed description of a vehicle-in-the-loop test method can be found in Berg~\cite{Berg.2014}.
Hallerbach~et~al.~\cite{Hallerbach.2018} call the concept of transferring perceived simulated objects to an automated driving function prototype-in-the-loop. In this approach, a real vehicle is continuously interacting with a traffic simulation while driving on a real-world proving ground.

\section{Example classifications}
\label{sec_examples}

In this section, the proposed method for classifying test benches and test bench configurations is evaluated by showing various examples. 
First, two example test benches are classified and, subsequently, two test bench configurations are derived and classified.

\subsection{Classification of test benches} 

For the classification of test benches, all elements which are provided at a specific test bench have to be taken into account. 
Therefor, the dimension environment sensor system has to be substantiated by the type of sensors available at a test bench because normally there is more than one type of environment sensors.

\begin{figure}[ht]
	\centering
	\includegraphics[width=0.8\columnwidth]{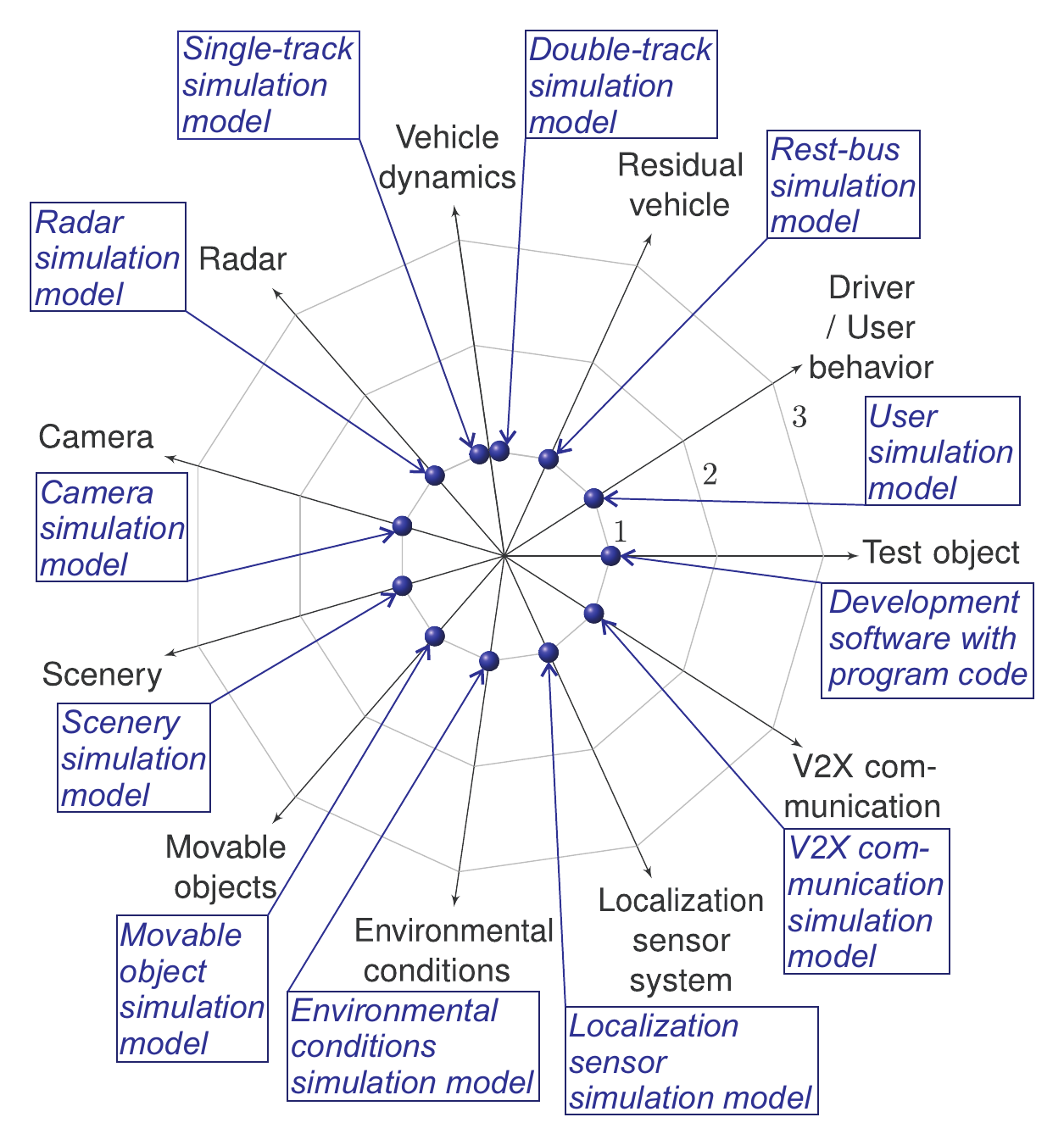}
	\caption{Classification of a specific software-in-the-loop test bench by a radar chart with the stages: 1~=~simulated, 2~=~emulated and 3~=~real (blue dots represent elements at this test bench)} 
	\label{fig_KV_SiL_Test_Bench}
\end{figure}

Fig.~\ref{fig_KV_SiL_Test_Bench} shows the classification of a specific software-in-the-loop test bench with all elements available at each dimension and stage.
This test bench provides two vehicle dynamics models: one single-track simulation model and one double-track simulation model. 
Therefore, we have two elements in the dimension vehicle dynamics at stage simulated. 
To emphasize these two simulation models, they are not drawn exactly on the arrow, but slightly offset.
Furthermore, there is one radar simulation model and one camera simulation model. 
Therefore, we substantiated the dimension environment sensor system by dimension radar and camera.
Each of these dimensions includes the corresponding simulation model.
For the sake of simplicity, this software-in-the-loop test bench provides one simulated element in all other dimension.

\begin{figure}[ht]
	\centering
	\includegraphics[width=0.8\columnwidth]{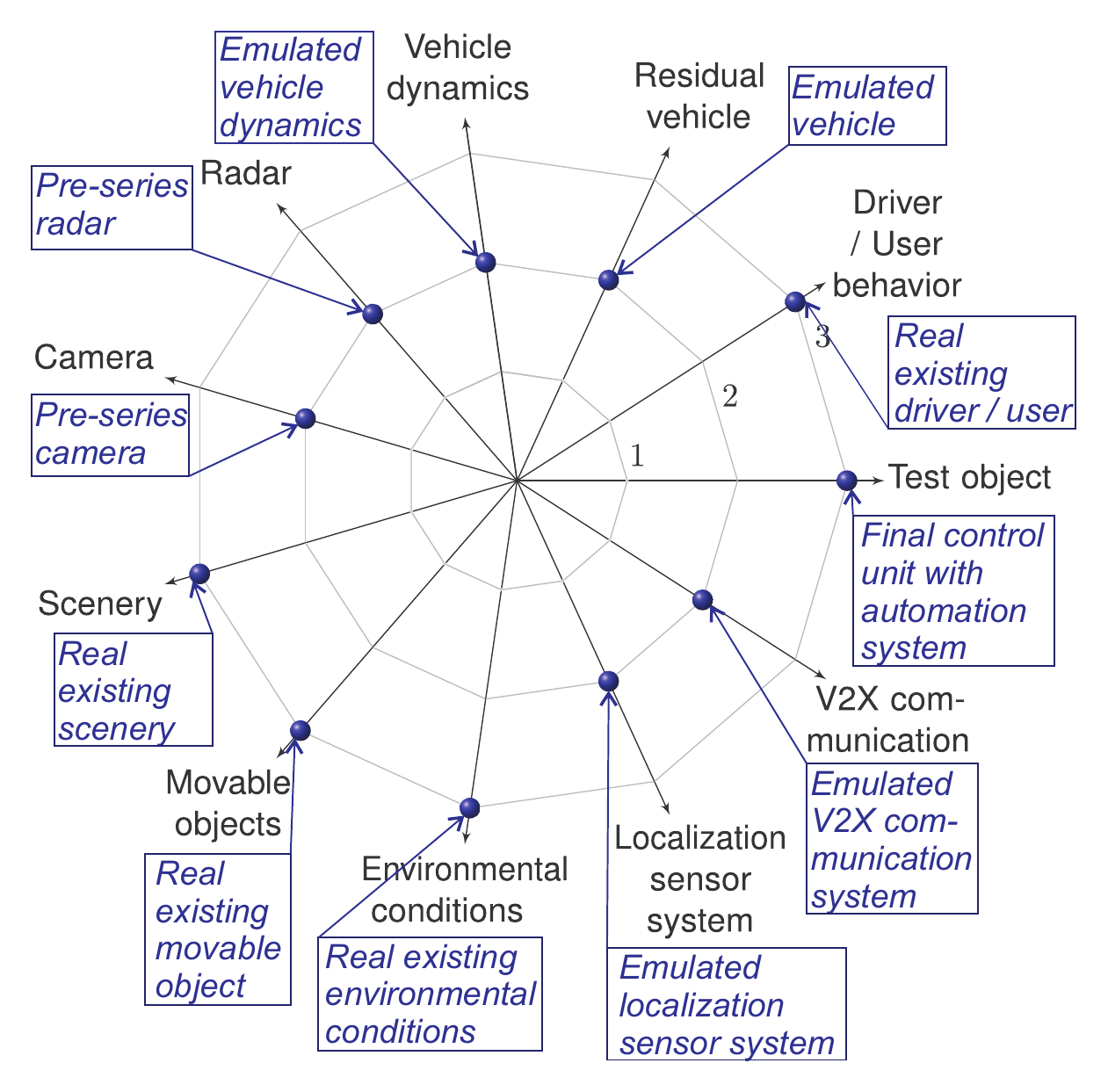}
	\caption{Classification of a specific test vehicle by a radar chart with the stages: 1~=~simulated, 2~=~emulated and 3~=~real (blue dots represent elements at this test bench)} 
	\label{fig_KV_TestVehicle}
\end{figure}

Fig.~\ref{fig_KV_TestVehicle} shows the classification of a specific test vehicle with all elements available in each dimension and stage.
For the sake of simplicity, this test vehicle provides one element in every dimension.

Every element shown in Fig.~\ref{fig_KV_SiL_Test_Bench} and Fig.~\ref{fig_KV_TestVehicle} has certain characteristics, such as validity, costs per operating time, and complexity in terms of necessary execution time.
Therefore, each test bench has certain properties. 
These characteristics are not part of this paper.

\subsection{Classification of test bench configurations}

Based on the classification of test benches, we have to derive test bench configurations in order to execute a test case.
For this, every possible composition of the available elements must be taken into account.

\begin{figure}[ht]
	\centering
	\includegraphics[width=0.78\columnwidth]{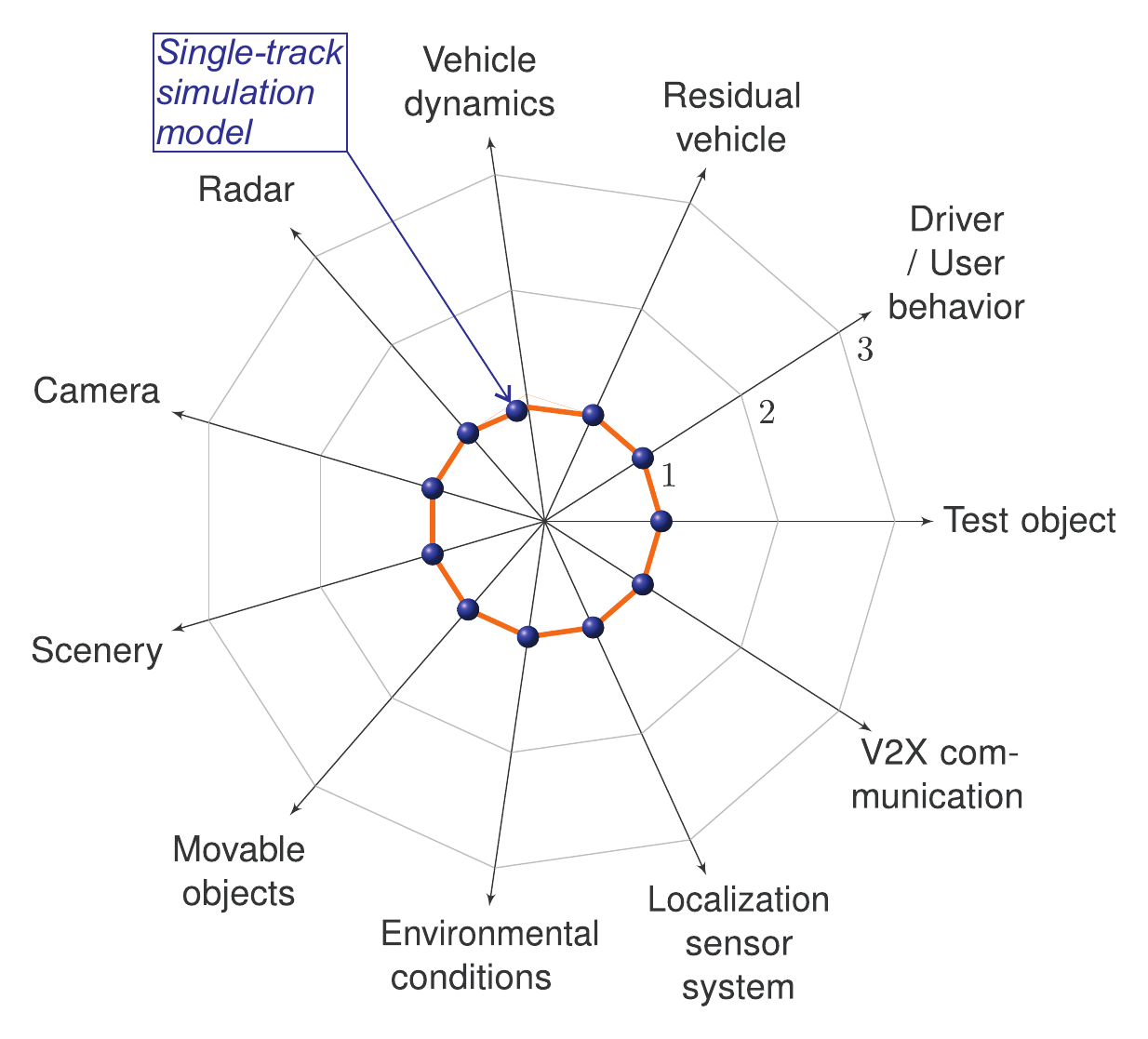}
	\caption{Classification of the test bench configuration 1 by a radar chart with the stages: 1~=~simulated, 2~=~emulated and 3~=~real (blue dots represent elements at this test bench configuration, the orange line visualizes the composition of elements)} 
	\label{fig_KV_Example_Test_Bench_Configuration_1}
\end{figure}

Based on the classification of the software-in-the-loop test bench shown in Fig.~\ref{fig_KV_SiL_Test_Bench}, we derive two test bench configurations. 
Fig.~\ref{fig_KV_Example_Test_Bench_Configuration_1} shows the classification of the test bench configuration 1, which contains the single-track simulation model. 
For the sake of clarity, the remaining elements are not listed. 
These elements are shown in Fig.~\ref{fig_KV_SiL_Test_Bench}.
The orange line visualizes the composition of the elements which are part of this test bench configuration.
The test bench configuration 2, which contains the double-track simulation model, can be classified the same way.

As a result, we have two different test bench configurations, each consisting of a different composition of elements.
In this case every test bench configuration has a different simulation model for the dimension vehicle dynamics.
The actual performance of the test bench is determined by the used test bench configuration.
Depending on the used test bench configuration the run of a specific test case may or may not provide valid test case results.
However, only if valid test case results are generated, can they be used for a safety validation. 

By this classification, it is thus possible to classify the properties of test bench configurations which serve as a basis for a systematic test case assignment.

\section{Conclusion \& Future Research}
\label{sec_conclusion}

In this paper, a method for the classification of test benches and test bench configurations was presented.
The classification is done by radar charts, where each dimension represents a specific functionality a test bench has to provide in order to execute a test case.
To derive these functionalities in a systematic way, we have analyzed the interfaces of a functional system architecture of an automated vehicle which we have connected with a driver or a user and the environment.
On this basis, dimensions for classifying test benches and test bench configurations were derived and a definition for each dimension was proposed.
For the evaluation of the classification method, we have illustrated selected examples of test benches and test bench configurations.

This classification method serves as a basis for a method to assign test cases to test bench configurations with regard to efficient and effective testing.
Currently, we are working on the test case assignment method.

\vspace{-0.5em}
\section*{Acknowledgment}

We would like to thank project members of the project PEGASUS funded by the German Federal Ministry for Economic Affairs and Energy for the productive discussions and the feedback on our approaches.
Our work is partially funded by the Robert Bosch GmbH and the Volkswagen AG.

\addtolength{\textheight}{-14.4cm}%
\vspace{-0.5em}
%
\bibliographystyle{IEEEtran}%
\bibliography{bib/references}%
\end{document}